\documentclass[11Pt]{article}
\usepackage{amssymb,euscript,latexsym,amsmath}
\usepackage{graphicx}
\usepackage[dvips]{epsfig}
\usepackage{color}
\usepackage{epstopdf}
\usepackage{setspace}
\usepackage{subfigure} 
\usepackage{cite}
\usepackage{hyperref}

\usepackage{caption2}

\DeclareGraphicsExtensions{.eps,.ps,.pdf}

\newcommand{\conj}[1]{\overline{#1}}
\newcommand{\real}[1]{\mathrm{Re} \left[ #1 \right]}

\newcommand{\eee}[1]{\mathrm{e}^{ #1 }}
\newcommand{\ii}{\mathrm{i}} 

\begin{document}

\title{\Large \bf  Pursuit and Synchronization in Hydrodynamic Dipoles} 
\author{\normalsize Eva Kanso and Alan Cheng Hou Tsang \\[2ex]
{\footnotesize Aerospace and Mechanical Engineering, University of Southern California} \\
{\footnotesize 854 Downey Way, Los Angeles, CA 90089-1191} \\
}

\maketitle

\begin{abstract}
We study theoretically the  behavior of a class of hydrodynamic dipoles. This study is motivated by recent experiments on synthetic and biological swimmers in microfluidic \textit{Hele-Shaw} type geometries. Under such confinement,  a swimmer's hydrodynamic signature is that of a potential source dipole, and the long-range interactions among swimmers are obtained from the superposition of dipole singularities. Here, we recall the equations governing the positions and orientations of interacting asymmetric swimmers in doubly-periodic domains, and focus on the dynamics of swimmer pairs. We obtain two families of `relative equilibria'-type solutions  that correspond to pursuit  and synchronization of the two swimmers, respectively. Interestingly, the pursuit mode is stable for large tail swimmers whereas the synchronization mode is stable for large head swimmers. These results have profound implications on the collective behavior reported in several recent studies on populations of confined  microswimmers.
\end{abstract}

\section{Introduction}

Active systems, i.e., systems driven internally by self-propelled individual units,  often exhibit rich collective behavior at the system's scale; a scale that is typically several orders of magnitude larger than the scale of the individual unit.  Such collective behavior naturally arises in disparate biological systems, from schools of fish~\cite{couzin:jtb2002a}  to suspensions of motile bacteria~\cite{cisneros:ef2007a} and assemblages of sub-cellular extracts~\cite{sanchez:n2012a}. 
It also emerges in inanimate systems such as driven and self-propelled droplets and reactive colloids~\cite{schaller:n2010a, bricard:n2013a}, and provide an attractive paradigm for reconfigurable smart materials~\cite{Richtering2006a} and biomedical devices~\cite{Jackson2013a}. 

The question of how these highly-coordinated collective motions arise from piecewise interactions among individual units has been the subject of intense research in the past few years. A well-studied example is the behavior of self-propelled  particles in a viscous fluid, \cite{saintillan:prl2008a}. Most of this work has focused on the instabilities and spatiotemporal fluctuations in three-dimensional (3D) systems. However, motivated by recent technological advances in producing and manipulating large ensembles of particles in microfluidic devices~\cite{bricard:n2013a, beatus:pr2012a, desreumaux:prl2013a}, attention began to shift to the collective dynamics of particles confined in quasi two-dimensional (2D) geometries. Geometric confinement changes drastically the nature of the hydrodynamic interactions among particles,~\cite{beatus:pr2012a}. The long-ranged hydrodynamic interactions in 3D are driven by the force dipoles exerted by self-propelled particles on the fluid medium,~\cite{saintillan:prl2008a}. In quasi-2D geometries, the solid walls  screen the force dipole contribution, making it subdominant in comparison with the potential dipole arising from incompressibility,~\cite{desreumaux:prl2013a}.  As a result, the long-range interactions among swimmers can be obtained from the superposition of dipole singularities. 

In this paper, we revisit the hydrodynamic dipole model proposed by Brotto \textit{et al.}~\cite{brotto:prl2013a} for asymmetric dumbbell swimmers in confined Hele-Shaw type geometries. The head-tail asymmetry causes a given swimmer to reorient, not only in response to the flow gradient as anticipated by Jeffery's equation,  but also in response to the flow velocity itself. This result is rooted in the fact that the lubrication forces between the swimmer and the solid walls hinder its advection by the fluid, inducing unequal translational motility coefficients at the swimmer's head and tail. In~\cite{brotto:prl2013a},  Brotto \textit{et al.} derived a kinetic theory-type model for a population of interacting swimmers and predicted a novel long-wave linear instability that leads to the emergence of  large-scale directed motion and polarization in isotropic populations of confined large head swimmers. Lefauve and Saintillan~\cite{lefauve:pre2014a} and Tsang and Kanso~\cite{tsang:pre2014a} used numerical simulations to explore the implications of these instabilities on the collective behavior in finite-sized populations of interacting swimmers.

The present paper  examines the detailed dynamics of a pair of asymmetric (head-tail) swimmers in doubly-periodic domains, where the orientation dynamics is dominated by the flow field itself, thus neglecting reorientation in response to the flow gradient as done in~\cite{brotto:prl2013a,lefauve:pre2014a}. We particularly focus on a special class of solutions where the two swimmers move with constant speed and at constant orientation. We find two families of these 
``equilibrium-like" solutions: (1) both swimmers swim side by side in a synchronized way; and (2) one swimmer tailgates the other. We analyze their stabilities and find that they depend on the details of the head-tail asymmetry. We conclude this work by discussing the significance of these results to the behavior of populations of swimmers.

Note that a dynamical theory of dipole interactions has also been pursued in two additional contexts. One motivation stems from the desire to obtain low-order representations of  two-dimensional, inviscid and  incompressible fluids in terms of  interacting particles such as point vortices and point dipoles, see, e.g.,~\cite{kulik:tmp2010a, smith:pnp2011a, smith:rcd2013a, newton:dcds2005a, yanovsky:pl2009a}. A shortcoming of these models is that 
the dipole's self-propelled speed is ill-defined; thus, a dipole, unless properly desingularized, induces 
infinite velocity through its center. 
Another motivation for dipole models that is closer to the focus of this paper grew out of efforts to examine the role of hydrodynamic coupling in fish schooling. It is a well-known result in fluid dynamics that the leading order flow of a self-propelled body is \textcolor{black}{that of a potential source dipole}. Kanso and co-workers proposed a finite dipole dynamical system that captures the far-field hydrodynamic interactions \textcolor{black}{of self-propelled bodies}~\cite{tchieu:prsa2012a, tsang:jnls2013a}. Each dipole consists of a pair of equal and opposite strength point vortices separated by a finite constant distance. By construction, the self-propelled speed is well-defined. \textcolor{black}{These finite dipoles are advected by the local flow and reorient in response to the local flow gradient. In particular, the finite dipole reorients according to velocity gradient in the direction transverse to the dipole orientation,} as opposed to reorienting in response to the flow gradient along the dipole's direction predicted by Jeffery's equation for slender bodies in viscous fluids.  \textcolor{black}{Intrigued by the similarities
between the finite dipole model~\cite{tchieu:prsa2012a, tsang:jnls2013a} and the dipole model of~\cite{desreumaux:epje2012a,brotto:prl2013a},} Kanso and Tsang~\cite{kanso:fdr2014a} presented a unified framework for deriving two point dipole models: a dipole consistent with the finite dipole model, appropriate for bluff bodies (fish) in potential flows, and another consistent with Jeffery's equation for slender bodies and equivalent to the microswimmer model employed in~\cite{desreumaux:epje2012a,brotto:prl2013a}. They further showed that, in unbounded domains, dipole pairs can synchronize their motions for a range of initial conditions; however, the details of the synchronized motion differ between the two 
models.

\textcolor{black}{The organization of this paper is as follows:  In Section~\ref{sec:model}, we formulate the equations of motions for a systems of dipoles in unbounded and in doubly-periodic domains. A detailed treatment of the dynamics of dipole pairs is conducted in Section~\ref{sec:results}. These results are discussed in Section~\ref{sec:discussion} in light of the large-scale simulations  performed on the same system in~\cite{lefauve:pre2014a,tsang:pre2014a}. }

\section{Problem Formulation}
\label{sec:model}

\paragraph{Microswimmer model.} Consider a microswimmer composed of two connected disks of radii $R_{tail}$ and $R_{head}$ located at $z_{tail}$ and $z_{head}$ respectively, where $z = x+ \ii y$  is the complex coordinate ($\ii = \sqrt{-1}$). Assume the two disks are connected by a frictionless rod of length $\ell$. The equations of motion for the swimmer's tail ($z_{tail}$) and head ($z_{head}$) can be written in complex notation as (see~\cite{brotto:prl2013a})
\begin{equation} 
\begin{split}
\label{eq:eomtail}
    \dot{\bar{z}}_{tail} & = U_o\eee{-\ii \alpha_o}+\mu_{tail}\bar{w}(z_{tail})+\lambda_{tail} \eee{-\ii \alpha_o},  \\[2ex]
    \dot{\bar{z}}_{head} & = U_o\eee{-\ii \alpha_o}+\mu_{head}\bar{w}(z_{head})-\lambda_{head} \eee{-\ii \alpha_o}. 
\end{split}
\end{equation}
Here, $U_o$ is the swimmer's self-propelled velocity, $\alpha_o$ its  orientation angle, and $w(z)$ is the velocity field of the ambient fluid. 
The bar notation denotes the complex conjugate, $\bar{z} = x - i y$. The coefficients $\mu_{tail}$, $\mu_{head}$ are the translational mobility coefficients whereas $\lambda_{tail}$, $\lambda_{head}$ are unknown Lagrange multipliers that enforce the constraint $| z_{head} - z_{tail}| = \ell$. In particular, the translational mobility coefficients $\mu_{tail}$ and $\mu_{head}$ arise from the balance of hydrodynamic drag and wall friction acting on the tail and head, and are decreasing functions of $R_{tail}$ and $R_{head}$ respectively, with values less than 1, see, e.g.~\cite{desreumaux:epje2012a, brotto:prl2013a}.

We define the hydrodynamic center of the swimmer to be $z_o = (\lambda_{tail}z_{head} + \lambda_{head}z_{tail})/(\lambda_{tail}+\lambda_{head})$.  Our goal is to rewrite the equations of motion~\eqref{eq:eomtail} 
 in terms of the swimmer's hydrodynamic center $z_o$ and orientation $\alpha_o$. Let $\ell \gg R_{tail}, R_{head}$, and use Taylor series expansion
to expand the flow velocity at the tail and head
\begin{equation} 
\begin{split}
\label{eq:velocitytail}
{\bar{w}}(z_{tail}) & = \bar{w}(z_o)+ \eee{\ii \alpha_o}\frac{\lambda_{tail}}{\lambda_{tail}+\lambda_{head}}  \left. \frac{d\bar{w}}{dz} \right|_{z_o}+\ldots \\ 
{\bar{w}}(z_{head}) & = \bar{w}(z_o)+\eee{\ii \alpha_o}\frac{\lambda_{head}}{\lambda_{tail}+\lambda_{head}} \left. \frac{d\bar{w} }{dz}\right|_{z_o}+\ldots .
\end{split}
\end{equation}
Substitute \eqref{eq:velocitytail} into~\eqref{eq:eomtail} 
to get that the equation governing the translational motion of the swimmer's center 
\begin{equation} 
\label{eq:eomcenter}
\begin{split}
\dot{\bar{z}}_o= U_o\eee{- \ii \alpha_o}+\mu \bar{w}(z_o) , 
\end{split}
\end{equation}
where $\mu  = {(\lambda_{head}\mu_{tail}+\lambda_{tail}\mu_{head})}/{(\lambda_{head}+\lambda_{tail})}$. 
To obtain the equation governing the rotational motion of the swimmer, note that, by definition, $\ell\eee{\ii \alpha_o}  = z_{head} - z_{tail}$, which gives, upon differentiating both sides with respect to time and further simplifications,
\begin{equation} 
\label{eq:eomori}
	 \dot{\alpha}_o  =  \real{\frac{(\dot{\bar{z}}_{head}- \dot{\bar{z}}_{tail}) \ii \eee{\ii \alpha_o}}{\ell}}.
\end{equation}
Here, Re denotes the real part of the expression in bracket. Now substitute \eqref{eq:eomtail} and \eqref{eq:velocitytail} into \eqref{eq:eomori} to get
\begin{equation} 
\label{eq:eomorientation}
	 \dot{\alpha}_o  =  \real{\nu_1 \frac{d \conj{w}}{dz} \ii \eee{2 \ii \alpha_o}+\nu_2  \conj{w} \ii  \eee{\ii \alpha_o}}.
\end{equation}
where $\dfrac{d \conj{w}}{dz} $ and $ \conj{w}$ are evaluated at $z_o$ and the constant parameters $\nu_1$ and $\nu_2$ are given by
\begin{equation} 
\label{eq:coeff}
\begin{split}
	\nu_1 = \frac{(\lambda_{head}\mu_{head}+\lambda_{tail}\mu_{tail})}{\lambda_{head}+\lambda_{tail}}, \qquad 
	\nu_2 = (\mu_{head}-\mu_{tail})/\ell.		
\end{split}
\end{equation}
The sign of $\nu_2$ dictates how the swimmer orients in local flow: it aligns to the local flow when $\nu_2>0$, that is, for large tail swimmers for which  $\mu_{head}-\mu_{tail}>0$ (because $\mu_{head/tail}$ is a decreasing function of $R_{head/tail}$,~\cite{beatus:pr2012a}),  and opposite to the local flow when $\nu_2 <0$ , that is, for large head swimmers for which $\mu_{head}-\mu_{tail}<0$.

\paragraph{Hydrodynamic interactions of multiple microswimmers.}  Consider the interaction of multiple microswimmers  in an unbounded fluid domain. 
By virtue of~\eqref{eq:eomcenter} and~\eqref{eq:eomorientation}, the dynamics of $N$ swimmers, all having the same self-propelled velocity $U$, can be expressed in concise complex notation
\begin{equation}
\label{eq:eom}
\begin{split}
  \dot{\conj{z}}_n  & = U \eee{-\ii \alpha_n}+ \mu \conj{w}(z_n), \\[1ex]
  \dot{\alpha}_n  & =  \real{\nu_1 \frac{d \conj{w}}{dz} \ii \eee{2 \ii \alpha_n}+\nu_2 \conj{w} \ii  \eee{\ii \alpha_n}}.
  \end{split}
\end{equation}
Here,  $z_n$ and $\alpha_n$ denote the position and orientation of each swimmer ($n=1,\ldots, N$).  To close the model, one needs to obtain an expression for the fluid velocity field $w(z)$. Recalling that each 
swimmer induces a far-field velocity which is that of a  potential source dipole 
\cite{beatus:pr2012a},  the far-field flow of a microswimmer  $j$ located at $z_j = x_j + \ii y_j$  and oriented at an arbitrary angle $\alpha_j$  can be described by the complex velocity $\conj{w}(z)=u_x-i u_y=\sigma \eee{\ii \alpha_j}/(z-z_j)^2$, where $\sigma$ is the dipole strength. Note that $\sigma=R^2 U$, where $R$ is the effective radius of the swimmer.  A microswimmer $n$ responds to the flow induced by all microswimmers  in the fluid domain, namely,
\begin{equation}
\label{eq:velocityDipole}
\conj{w}(z_n)= \sum_{\stackrel{j\neq n}{j=1}}^N \sigma  \dfrac{\eee{\ii \alpha_j}}{(z_n-z_j)^2}.
\end{equation}

\paragraph{Microswimmers in doubly-periodic domains.} When the swimmers are placed in a doubly-periodic domain, one needs to take into account, not only the velocity field induced by the swimmers themselves but also the effect of their image system. A given swimmer $n$ has a doubly-infinite set of images. Thus, evaluating $w(z)$ requires the evaluation of conditionally-convergent, doubly-infinite sums of terms that decay as $1/|z|^2$. These sums are evaluated using an approximate numerical approach  in~\cite{lefauve:pre2014a}. In~\cite{tsang:pre2014a}, we offered a closed-form analytic expression for these infinite sums in terms of the Weierstrass elliptic function, namely,
\begin{equation}
	\label{eq:velocityDipolePeriodic}
	 \conj{w}(z) = \sum_{n=1}^N\sigma \rho( z-z_n;\omega_1,\omega_2)\eee{\ii \alpha_n}.
\end{equation}
The Weierstrass elliptic function $\rho(z)$ is given by $\rho\left(z;\omega_{1},\omega_{2}\right)=\frac{1}{z^2}+\sum_{k,l} \left(\frac{1}{(z-\Omega_{kl})^2} -\frac{1}{\Omega_{kl}^{2}}\right)$, with $\Omega_{kl}=2k\omega_{1}+2l\omega_{2}$,  $k,l\in \mathbb{Z} \!-\!\{0\}$,
and $\omega_{1}$ and  $\omega_{2}$ being the half-periods of the doubly-periodic domain. This function has infinite numbers of double pole singularities located at $z=0$ and $z=\Omega_{kl}$, corresponding to the $1/|z|^2$ singularities induced by the potential dipoles. Equations \eqref{eq:eom} and \eqref{eq:velocityDipolePeriodic} form a closed system for $N$ swimmers in a doubly-periodic domain.

We conclude by writing the system of equations \eqref{eq:eom} and \eqref{eq:velocityDipolePeriodic} in dimensionless form using the swimmers radius $R$ as a length scale  and  $R/U$ as a time scale. That is, we introduce the dimensionless spatial variable $\tilde{z}=z/R$ and time variable $\tilde{t}=t U/R$. We then drop the tilde notation assuming all variables are non-dimensional. Equations \eqref{eq:eom} and \eqref{eq:velocityDipolePeriodic} have the same form but  the parameters $U$ and $\sigma$ are now normalized to one, that is, $U=1$ and $\sigma=1$. The parameter values $\mu$, $\nu_1$ and $\nu_2$ are also non-dimensional.

\section{Pursuit and synchronization}
\label{sec:results}

We consider two microswimmers in a doubly-periodic domain, and focus on their dynamic response when $\nu_1=0$, that is, when their alignment with the flow gradient is negligible. In this case, the orientation dynamics is dominated by alignment with the flow due to head-tail hydrodynamic asymmetry. 

\begin{figure}[!t]
\begin{center}
\includegraphics[width=\textwidth]{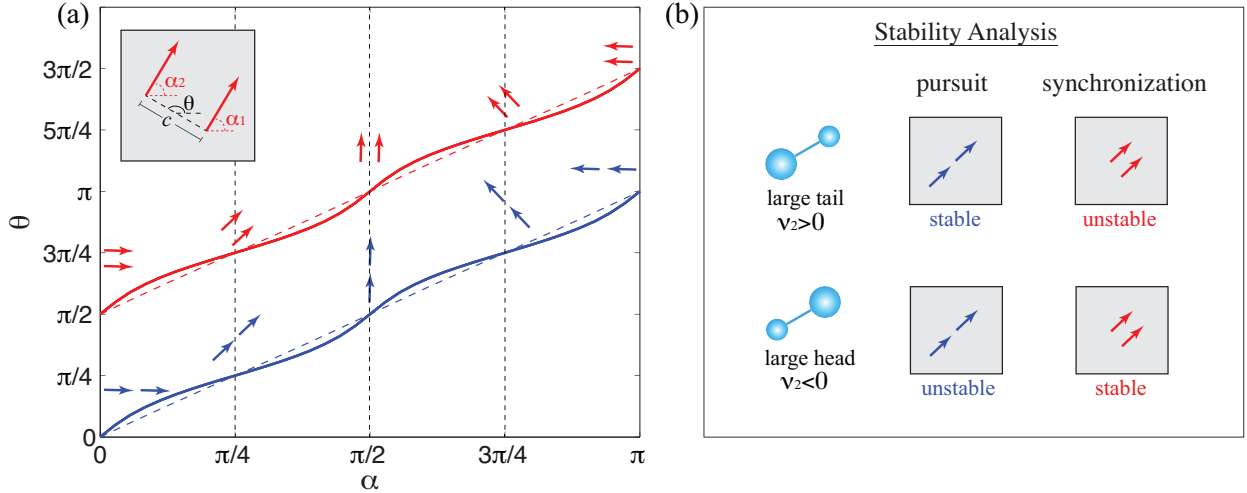}
\caption{ (a) Two modes of  solutions are obtained from~\eqref{eq:condition}:  a pursuit mode (blue) where the swimmers trail one another and a synchronization mode (red) where they swim side by side. The shown solid curves correspond to  $\omega_1=-\ii \omega_2=5$ and the dashed lines to $\omega_1=-\ii \omega_2=10$. The separation distance $c$ is set to $c=4$. (b) Summary of the stability analysis for these two modes.}
 \label{fig:twodipolesratio}
\end{center}
\end{figure}

\paragraph{Periodic solutions and relative equilibria.} We look for special solutions where the two swimmers move at the same  velocity and orientation for all time. That is, we look for solutions where $\dot{z}_1 = \dot{z}_2=$ constant and $\dot{\alpha}_1 = \dot{\alpha}_2 = 0$. To obtain the initial conditions that lead to this behavior, it is convenient to rewrite the equations of motion~(\ref{eq:eom},\ref{eq:velocityDipolePeriodic}) in terms of the reduced coordinate $z_1-z_2$ which we set to  $z_1-z_2= \beta = c \eee{\ii \theta}$ (see inset of Figure~\ref{fig:twodipolesratio}(a)). To this end, one gets 
\begin{equation} 
\begin{split}
\label{eq:eomtwodipoles}
\dot{\bar{\beta}} & = \eee{- \ii \alpha_1} -  \eee{- \ii \alpha_2}+\mu \rho(\beta) ( \eee{ \ii \alpha_2} - \eee{ \ii \alpha_1}), \\[2ex]
\dot{\alpha}_1 & = \dot{\alpha}_2= \nu_2 \text{Re}[\ii \eee{\ii (\alpha_1+\alpha_2)} \rho(\beta)].
\end{split}
\end{equation}
The translation equation for $\dot{\beta}$ is identically zero when $\alpha_1 =\alpha_2=\alpha$. Whereas to guarantee $\dot{\alpha}_1 = \dot{\alpha}_2 = 0$, one must satisfy the condition
\begin{equation} 
\begin{split}
\label{eq:condition}
	\left\{ \begin{array}{l} 
	{\text{Re}[\rho(c  \eee{- \ii \theta})]}=0 \textrm{ \ therefore \ }  \alpha = \dfrac{\pi}{4},  \dfrac{3\pi}{4}  \textrm{\ \ and \ }  (\alpha,\theta) = \left\{ \begin{array}{l}   (\alpha,\alpha) \\ 
	(\alpha,\alpha+ \pi/2) \end{array} \right. \\
	\textrm{or} \\[2ex] 	
	\dfrac{\text{Im}[\rho(c  \eee{- \ii \theta})]}{\text{Re}[\rho(c  \eee{- \ii \theta})]} = -\tan{2 \alpha}, \qquad {\text{Re}[\rho(c  \eee{- \ii \theta})]} \neq 0, 
		\end{array} \right. 
\end{split}
\end{equation}
\textcolor{black}{A total of ten strict relative equilibria of the two swimmers are depicted schematically in Figure \ref{fig:twodipolesratio}(a)}. Namely, the five solutions given by $\alpha = 0, \dfrac{\pi}{4}, \dfrac{\pi}{2}, \dfrac{3\pi}{4}, \pi$ and  $\theta = \alpha$  correspond to the two dipoles moving  parallel to each other in a   ``pursuit" mode (blue arrows), whereas the five solutions given by $\alpha = 0, \dfrac{\pi}{4}, \dfrac{\pi}{2}, \dfrac{3\pi}{4}, \pi$ and  $\theta = \alpha + \pi/2$ correspond to the two dipoles moving  in tandem in  a ``synchronized" mode (red arrows).

\begin{figure}[!t]
\begin{center}
\includegraphics[width=0.75\textwidth]{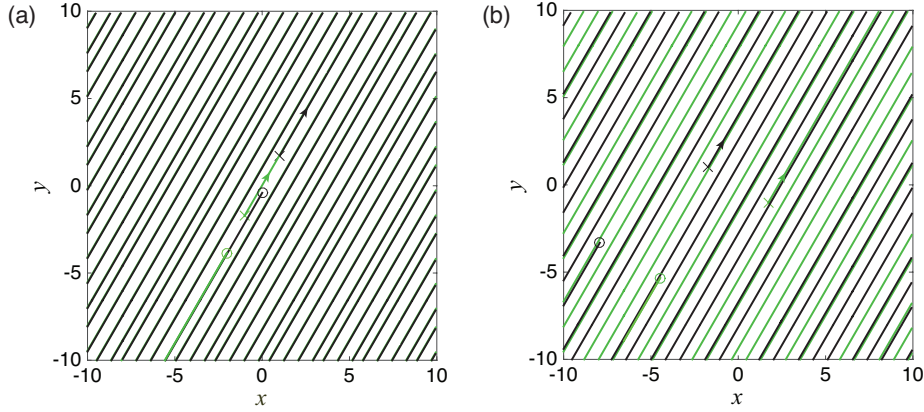}
\caption{Aperiodic behavior of two dipoles in doubly-periodic domain. The parameter values are $\alpha=\pi/3$, $z_1(0)=-2\exp(i\theta)$, $z_2(0)=2\exp(i\theta)$, while $\theta$ is obtained by numerically solving the first condition in \eqref{eq:condition}. The positions of the dipoles are marked by `$\times$' at $t=0$ and by `$o$' at the end of the integration.
As time progresses, the two trajectories densely fill the whole domain.}
 \label{fig:aperiodic}
\end{center}
\end{figure}

\begin{figure}[!t]
\begin{center}
\includegraphics[width=0.75\textwidth]{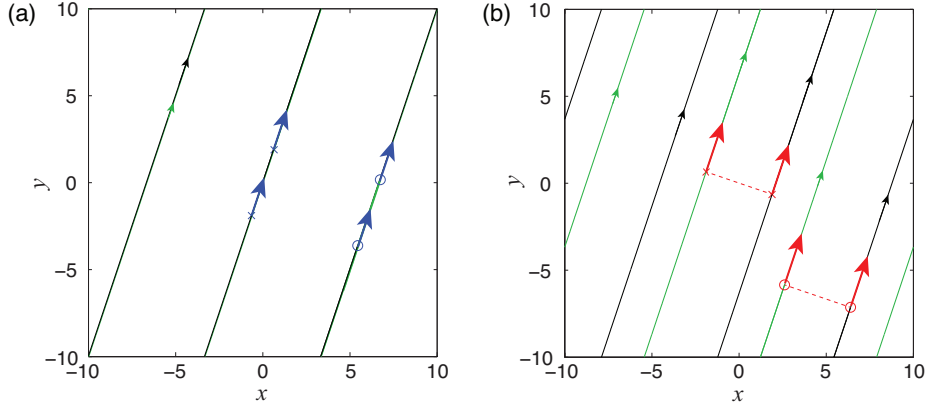}
\caption{ (a) Pursuit and (b) synchronization in two dipoles undergoing periodic motion. The parameter values are $\alpha=\tan^{-1}(3)$, $z_1(0)=-2\exp(i\theta)$, $z_2(0)=2\exp(i\theta)$, while $\theta$ is obtained by numerically solving the first condition in \eqref{eq:condition}. The positions of the dipoles are marked by `$\times$' at $t=0$ and by `$o$' at the end of the integration time $t = 80$.}
 \label{fig:twodipolesperiodic}
\end{center}
\end{figure}

The second set of solutions, that is, the values of $(\alpha, \theta)$ for which ${\text{Im}[\rho(c  \eee{- \ii \theta})]}/{\text{Re}[\rho(c  \eee{- \ii \theta})]} = -\tan{2 \alpha}$ and  ${\text{Re}[\rho(c  \eee{- \ii \theta})]} \neq 0$, are not analytically available and need to be computed numerically. Figure \ref{fig:twodipolesratio}(a) shows the values of $(\alpha,\theta)$ that satisfy these conditions -- clearly, two branches of solutions are obtained. These solutions depend implicitly on the domain size $(\omega_1,\omega_2)$ and on $c$, the separation distance between the two swimmers. In other words, for a choice of domain size and separation distance $c$, $(\alpha,\theta)$ are computed accordingly such that the two dipoles move at the same constant velocity and orientation for all time. 
The two branches shown in Figure \ref{fig:twodipolesratio}(a) correspond to two modes of behavior: a pursuit mode where one swimmer trails the other, and a synchronization mode where the two swimmers move side by side.
These solutions,  while they correspond to the dipoles moving at constant velocity and orientation, can exhibit 
two distinct types of dynamical behavior due to the doubly-periodic nature of the domain, namely, they could lead to aperiodic and periodic motion of the dipoles. Aperiodic motion refers to the case where the  paths of the dipoles densely fill the whole domain, \textcolor{black}{as shown in Figure \ref{fig:aperiodic}}. This seems to be the generic behavior for arbitrary initial conditions. Periodic behavior refers to trajectories that satisfy the condition 
\begin{equation} 
\label{eq:onedipoleperiodic}
\bar{z}_1(T)= \bar{z}_1(0) + 2 p \omega_1 + 2 q \omega_2, \qquad 
\bar{z}_2(T)= \bar{z}_2(0) + 2 p \omega_1 + 2 q \omega_2
\end{equation}
where $p$ and $q$ are integers and $T$ is the period of the motion. This amount to the additional condition
\begin{equation} 
\label{eq:onedipoleperiodic2}
 \alpha_1(0)=  \alpha_2(0) = \tan^{-1}(\frac{q}{p}).
\end{equation}
The ratio of $q/p$ indicates the ratio of the number of times the dipole crosses the $y$ and $x$ axes in one period $T$.   Figure~\ref{fig:twodipolesperiodic} depicts the periodic behavior of two dipoles in pursuit and synchronization modes for $q/p = 3$.


\begin{figure}[!t]
\begin{center}
\includegraphics[width=0.75\textwidth]{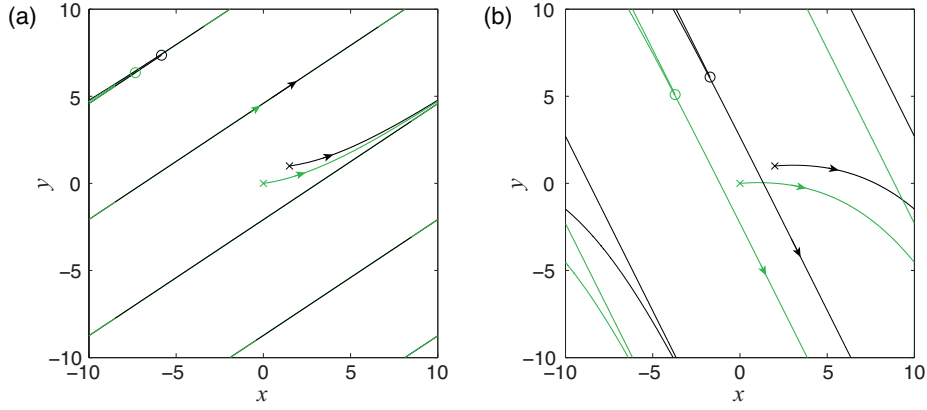}
\caption{Pursuit and synchronization modes as attracting modes. Two dipoles hone in on quasi periodic trajectories where: (a)  one dipole is in pursuit of the other for  $\nu_2=0.5$. (b) the two dipoles synchronize and move along parallel trajectories for $\nu_2=-0.5$. The initial conditions are $z_1(0)=\alpha_1(0)=\alpha_2(0)=0$ while $z_2(0)=1.5+1\ii$ in (a) and  $z_2(0)=2+1\ii$ in (b).}
 \label{fig:attractor}
\end{center}
\end{figure}

\paragraph{Stability analysis.} We analyze the linear stability of the pursuit and synchronization modes by considering small perturbations $\delta \beta = \delta \beta_x + \ii \delta \beta_y$, $\delta \alpha_1$ and $\delta \alpha_2$  about  $\beta =  c \eee{\ii \theta}$ ($\beta_x = c \cos\theta$, and $\beta_y= c \sin\theta$) and $\alpha_1 = \alpha_2 = \alpha$, with $(\alpha,\theta)$ satisfying~\eqref{eq:condition}.  We linearize equations~\eqref{eq:eomtwodipoles} accordingly.  The linearized equations can be written in matrix form as follows:
\begin{equation}
\label{eq:perturbtwodipoles}
\dfrac{d}{dt}\left(\begin{array}{c} \delta \beta_x \\ \delta \beta_y \\ \delta \alpha_1 \\ \delta \alpha_2 \end{array}\right) = M \left(\begin{array}{c} \delta \beta_x \\ \delta \beta_y \\ \delta \alpha_1 \\ \delta \alpha_2 \end{array} \right), 
\end{equation}
where the Jacobian matrix $M$ is given by
\begin{equation}
\label{eq:jacobian}
M =  \left(\begin{array}{cccc} 0 & 0 & -\sin\alpha-\mu\text{Re}[\ii \eee{\ii \alpha} \rho(\beta)] & \sin\alpha+\mu\text{Re}[\ii \eee{\ii \alpha} \rho(\beta)] \\
0 & 0 & -\cos\alpha-\mu\text{Im}[\ii \eee{\ii \alpha} \rho(\beta)] & \cos\alpha+\mu\text{Im}[\ii \eee{\ii \alpha} \rho(\beta)]\\ 
\nu_2 \text{Re}[\ii \eee{\ii 2 \alpha} \rho^{\prime}(\beta)] & -\nu_2 \text{Re}[\eee{\ii 2 \alpha} \rho^{\prime}(\beta)] 
& -\nu_2 \text{Re}[\eee{\ii 2 \alpha} \rho(\beta)] & -\nu_2 \text{Re}[ \eee{\ii 2 \alpha} \rho(\beta)] \\
\nu_2 \text{Re}[\ii \eee{\ii 2 \alpha} \rho^{\prime}(\beta)] & -\nu_2 \text{Re}[\eee{\ii 2 \alpha} \rho^{\prime}(\beta)] 
& -\nu_2 \text{Re}[\eee{\ii 2 \alpha} \rho(\beta)] & -\nu_2 \text{Re}[ \eee{\ii 2 \alpha} \rho(\beta)]\end{array} \right).
\end{equation}
We compute the eigenvalues numerically and find that, for large tail swimmers $\nu_2>0$, the pursuit mode is stable, whereas for large head swimmers $\nu_2<0$, the synchronization mode is stable. Our findings are summarized in Figure~\ref{fig:twodipolesratio}(b).

We test our results numerically by integrating the nonlinear equations~(\ref{eq:eom}, \ref{eq:velocityDipolePeriodic}) for arbitrary choices of  initial conditions. Interestingly, the pursuit and synchronization modes seem to be globally attracting modes in the case of large tail and large head swimmers, respectively. Figure~\ref{fig:attractor}(a)  shows a depiction of two large-tail swimmers honing in on quasi-periodic pursuit trajectories, while (b) depicts two large-head swimmers synchronizing their motion in finite time to swim side by side.

\paragraph{The limit of  unbounded domain.} We conclude this section by noting that in the limit of infinite domain, the solutions~\eqref{eq:condition} of the doubly-periodic system~(\ref{eq:eom}, \ref{eq:velocityDipolePeriodic}) converge to the relative equilibria  of the unbounded system~(\ref{eq:eom}, \ref{eq:velocityDipole}).  In the unbounded system, the relative equilibria can be obtained either by symmetry arguments or by analytical manipulation of the equations of motion. Namely, one has two families of relative equilibria $\theta=\alpha$ and $\theta=\alpha+\pi/2$ which correspond to pursuit and synchronization trajectories, respectively.  The convergence of the solutions in~\eqref{eq:condition} to these solutions is relatively fast, as indicated in Figure~\ref{fig:twodipolesratio}(a).
As $\omega_1$, $\omega_2 \rightarrow \infty$, the Jacobian matrix $M$ converges to 
\begin{equation}
\label{eq:jacobian2}
M_{\infty} = \left(\begin{array}{cccc} 0 & 0 & -\sin\alpha+\dfrac{\mu}{c^2} \sin(\alpha-2 \theta) & \sin\alpha-\dfrac{\mu}{c^2} \sin(\alpha-2 \theta)\\[1ex]
0 & 0 & -\cos\alpha- \dfrac{\mu}{c^2} \cos(\alpha-2 \theta) & \cos\alpha+ \dfrac{\mu}{c^2} \cos(\alpha-2 \theta)\\[1ex]
  \dfrac{2\nu_2}{c^3} \sin(2\alpha-3 \theta)& \dfrac{2 \nu_2}{c^3}\cos(2\alpha-3 \theta)
& - \dfrac{\nu_2}{c^2}  \cos(2\alpha-2 \theta) & - \dfrac{\nu_2}{c^2} \cos(2\alpha-2 \theta) \\[1ex]
\dfrac{2\nu_2}{c^3} \sin(2\alpha-3 \theta) & \dfrac{2\nu_2}{c^3} \cos(2\alpha-3 \theta)
& - \dfrac{\nu_2}{c^2} \cos(2\alpha-2 \theta) & - \dfrac{\nu_2}{c^2} \cos(2\alpha-2 \theta) \end{array} \right).
\end{equation}
The corresponding eigenvalues are $[0$, $0$, $0$, $\mp 2  \nu_2 /c^2]$. The eigenvalue $-2\nu_2/c^2$ corresponds to the pursuit mode where $\alpha = \theta$, whereas $+2\nu_2/c^2$ corresponds to the synchronization mode. This means that, for large tail swimmers with $\nu_2>0$, the pursuit mode is linearly stable and the synchronization mode is unstable, whereas  for large head swimmers  when $\nu_2>0$, the opposite is true, thus confirming the results obtained above for finite-sized doubly-periodic domains.

\section{Discussion}
\label{sec:discussion}

\begin{figure}[!t]
\begin{center}
\includegraphics[width=0.95\textwidth]{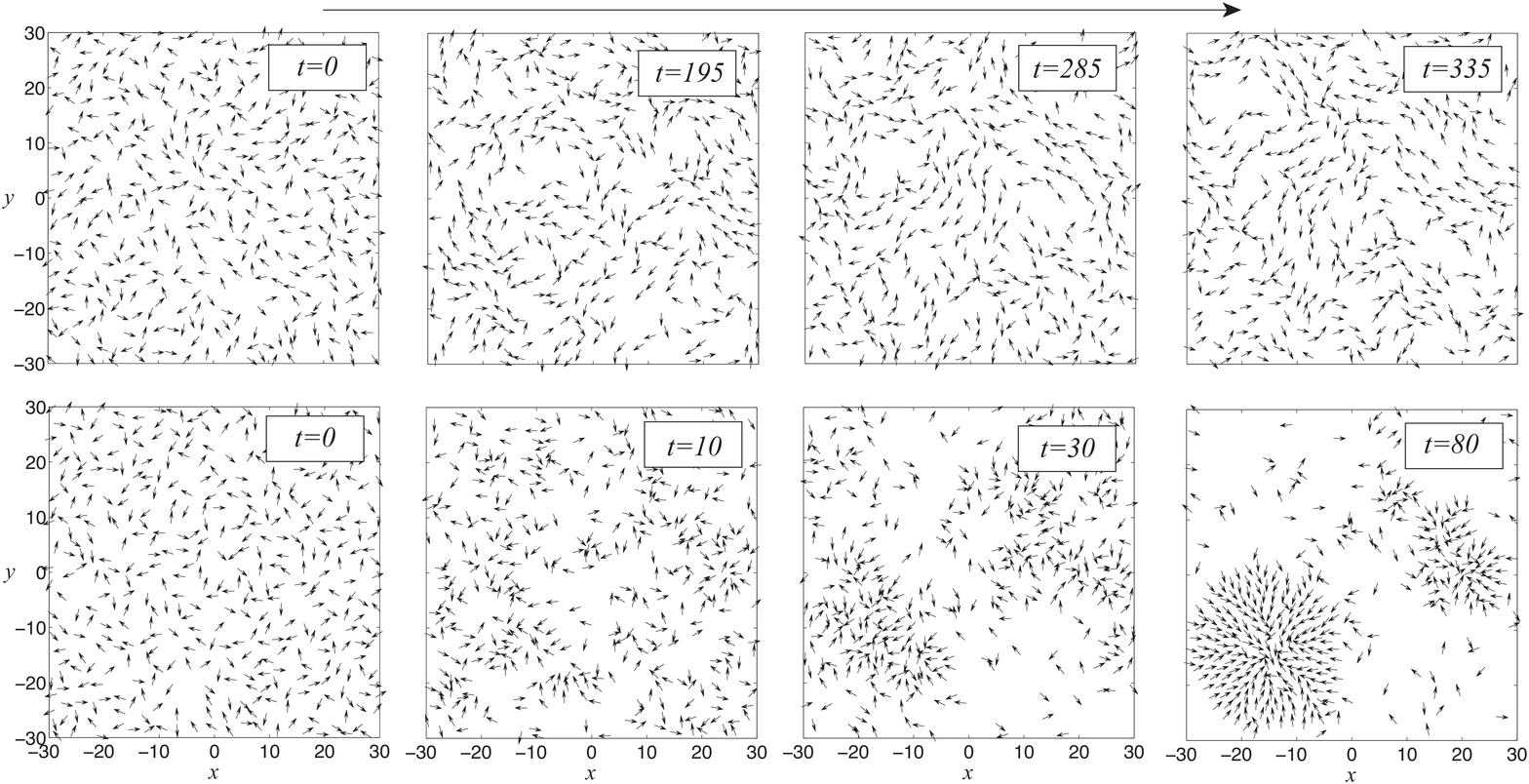}
\caption{Emergent collective behavior in large tail swimmers (top row) and large head swimmers (bottom row) starting from a uniform isotropic distribution. Large tail swimmers tend to tailgate each other, thus forming active lanes, while large head swimmers tend to form stationary clusters. Parameter values are $\mu = 0.9$, $\nu_2 = 1$ (top row) and $\nu_2=-0.5$ (bottom row) in a total population of $400$ swimmers.}
 \label{fig:populations}
\end{center}
\end{figure}

We revisited the hydrodynamic dipole model governing the interaction of asymmetric microswimmers in Hele-Shaw confinement,~\cite{brotto:prl2013a}.  Following~\cite{tsang:jnls2013a,tsang:pre2014a}, we obtained a closed-form expression for the velocity field induced by the swimmers and their image system in doubly-periodic domains. We treated in details the dynamics of two interacting swimmers, and found two special solutions that correspond to pursuit and synchronization of the two dipoles. The pursuit mode is stable and attracting for large tail swimmers while the synchronization mode is stable and attracting for large head swimmers. By attracting, we mean that, starting from arbitrary initial conditions,  large tail swimmers tend to tailgate each other while large head swimmers tend to synchronize their motion in finite time to swim side by side. These results are particularly interesting in light of the collective behavior reported in~\cite{lefauve:pre2014a,tsang:pre2014a} on populations of such swimmers. 
In these works, large tail swimmers  were observed to ``develop active lanes"~\cite{lefauve:pre2014a}  and ``tail-gate each other"~\cite{tsang:pre2014a}, as shown in Figure~\ref{fig:populations}(top row), which suggests that the pursuit mode remains stable as the system size increases.  \textcolor{black}{Note that, to generate Figure~\ref{fig:populations}, we integrate equations~\eqref{eq:eom} and~\eqref{eq:velocityDipolePeriodic} for a population of $400$ dipoles, starting from a uniform isotropic distribution and using the parameter values 
$\mu = 0.9$, $\nu_1 = 0$ and $\nu_2 = 1$ (large tail) or $\nu_2=-0.5$ (large head).}

Populations of large head swimmers were shown to form heavily polarized sharp density waves in~\cite{lefauve:pre2014a}, consistent with predictions based on linear stability analysis of a kinetic-type continuum model~\cite{brotto:prl2013a}. One could conjecture that the synchronization mode observed here in pairs of large head swimmers may be responsible for the global polarization observed in~\cite{lefauve:pre2014a}. However, this thinking is too simplistic. The emergence of global polarization patterns in finite size populations is not intuitive given the nature of dipolar interactions among the swimmers. Further, these polarized density waves were not observed in the detailed parametric study reported in~\cite[Figure 7]{tsang:pre2014a}.  Instead,~\cite{tsang:pre2014a} reported, in agreement with unpublished results by Levaufe and Saintillan, that large head swimmers tend to form stationary clusters  (see Figure~\ref{fig:populations}(bottom row)), which are  not predicted by the linear stability analysis of~\cite{brotto:prl2013a}.  
All this is to say that the global patterns of the finite size systems in \cite{lefauve:pre2014a} and \cite{tsang:pre2014a} are in agreement, except for the global polarization pattern.   This inconsistency may be due to differences in the system size -- thousands of particles in~\cite{lefauve:pre2014a}  versus hundreds in Figure~\ref{fig:populations} and in~\cite{tsang:pre2014a} --  or to differences in the details of the numerical implementation. In  \cite{lefauve:pre2014a}, the  point dipole model is desingularized and hydrodynamic interactions are approximated for fast computations, whereas \cite{tsang:pre2014a} use a local repulsion potential for collision avoidance and accurately account for hydrodynamic interactions and the doubly-infinite image system. \textcolor{black}{While the difference in numerical implementation may play a role, the system size may be the main reason 
why polarized waves are not observed in~\cite{tsang:pre2014a}. Brotto et al.~\cite{brotto:prl2013a} predicted this behavior for a continuous kinetic-like model, therefore it is not surprising that it is not reproduced by a fully nonlinear model with only a few hundred swimmers.}
Irrespective of the reason, the results reported in this study suggest that the global polarization mode in large head swimmers is not ``robust" to system perturbances, whereas the pursuit mode in large tail swimmers is.


\bibliographystyle{unsrt}
\bibliography{reference_arXiv}

\end{document}